# Knowledge Phenomenology Research of Future Industrial Iconic Product Innovation

Xu Jiang*, Qu Haoxiang

(College of Design and Innovation, Tongji University, Shanghai 200092, China)

**Abstract:** Iconic products, as innovative carriers supporting the development of future industries, are key breakthrough points for driving the transformation of new quality productive forces. Promoting innovation in iconic products requires adherence to philosophical guidance, abandoning reverse thinking characterized by imitation and following trends, grasping the fundamental principles of innovative design, and enhancing the capacity for independent innovation in major technological equipment. This article is grounded in the philosophy of technology and examines the evolution of human civilization to accurately identify the patterns of product innovation. By integrating theories from systems science, it analyzes the intrinsic logical differences between traditional products and iconic products. The study finds that iconic products are based on a comprehensive knowledge system that integrates explicit and tacit knowledge, enabling them to adapt to complex dynamic environments. Therefore, based on the method of phenomenological essence reduction and the process of specialized knowledge acquisition, this study establishes the first principle of knowledge phenomenology: "knowledge generation—moving from the tacit to the explicit—moving from the explicit to the tacit—fusion of the explicit and tacit." Grounded in knowledge phenomenology, it reconstructs the product design evolution process and establishes a forward innovative design framework for iconic products, consisting of "design problem space—explicit knowledge space—tacit knowledge space—innovative solution space." Furthermore, based on FBS design theory, it develops a disruptive technology innovation forecasting framework of "technology problem space—knowledge base prediction—application scenario prediction—coupled technology prediction," which collectively advances the innovation systems engineering of iconic products. In light of the analysis of the global future industrial competitive landscape, it proposes a strategy for enhancing embodied intelligence in iconic products. To efficiently promote the innovation of iconic products, it is essential to establish and improve forward innovative design methods in specialized fields, forecast disruptive technologies, categorize and formulate technology roadmaps, and strengthen the industrial application of technological products.

## 1 Introduction

The rapid advancement of next-generation artificial intelligence technology has emerged as a potent driver propelling the Fourth Industrial Revolution. By generating new resources, products, and business models, it facilitates the intelligent transformation of traditional industries, instigates transformative changes, and leads humanity into the era of generalized strong artificial intelligence[1-2]. When science and technology flourish, the nation prospers; when science and technology are robust, the country becomes powerful[3]. How to seize the strategic opportunities presented by the new wave of technological revolution in the age of artificial intelligence is a critical contemporary challenge for promoting the high-quality development of the national economy. Developed nations have established forward-looking, high-end equipment products as priority development areas to expedite the incubation of future industries. For instance, the United States, through its "Endless Frontier Act," has emphasized the advancement of artificial intelligence, cutting-edge computer hardware and software, robotics, and other future-oriented technology products. Japan's Blueprint for a New Industrial Structure outlines advanced technology products for "Society 5.0," including autonomous vehicles, innovative pharmaceuticals, and advanced material manufacturing. France prioritizes sustainable agricultural equipment, biofuels, and future telecommunications networks as key pillars of its "acceleration strategy." Germany is fostering future industries by investing in quantum computing, 6G communications, and electric vehicles[4]. In this context, China has taken the initiative to enhance strategic

**Corresponding Author:** *Xu Jiang is a professor from the College of Design and Innovation, Tongji University. His major research fields include innovative design strategy, design engineering and computing. E-mail: xujzju@163.com

**Funding Project:** National Key R&D Program of China (2022YFB3303300); Chinese Academy of Engineering project "Research on the Development Strategy of Jiangxi's Innovative Design Industry in the Age of Digital Intelligence" (2022-03JXZT-05) deployment by clearly defining development directions for six major industries, including future manufacturing, future information technology, and future space exploration. This has resulted in the creation of iconic products such as humanoid robots, quantum computers, and brain-computer interfaces[5]. The future industry is the transformation of disruptive technological innovation into real productive forces, with remarkable uncertainty and instability, resulting in extremely high risks and difficulties in cultivating new technologies and products. Therefore, it is necessary to start from the deep philosophical theory, grasp the law of product innovation and development, scientifically plan the development path of the future industry, and forge new-quality productive forces.

Iconic products are a group of key products based on the field of major technical equipment research and aimed at meeting the major strategic demands of the country and the needs for a better life of the people. They are designed to build a world-leading technical equipment system and create new competitive advantages for future industries. The related research is mainly divided into the following three aspects: At the strategic level, based on the perspectives of enterprise, national competitiveness and high-quality development[6-8], the establishment of a management decision-making system for the product innovation process is explored[9-10]; At the methodological level, based on the common characteristics of product innovation activities and rational thinking patterns, the scientific, axiomatic and procedural theories and methods of product innovative design are explored[11-13]; At the practical level, for major engineering equipment and product design objects, cutting-edge theories and technical methods are integrated to support product innovative design[14-15].

Based on the existing research findings, this paper aims to further deepen the analysis at the levels of philosophical thought and theoretical methodology. Specifically, in response to the significant strategic requirements for developing future industries and high-quality productive forces, and by integrating the evolution laws of human product innovation, this study elucidates the internal logic and potential challenges associated with iconic product innovation. It synthesizes the epistemology of phenomenology and philosophy of technology to establish a foundational principle of knowledge phenomenology for iconic product innovation. Building on this knowledge phenomenology, the paper explores the forward-looking innovative design architecture for iconic products, reverse-engineers and forecasts disruptive technological innovations, and establishes a system engineering framework for iconic product innovation. Considering the global competitive landscape of future industries, it proposes an embodied intelligence enhancement strategy to provide theoretical references for related research.

## 2 Innovations in Iconic products Lead the Development of New-Quality Productive Forces in the Country

### 2.1 Technological Philosophy Identification of Human Product Innovation Evolution Patterns

From agricultural civilization to modern industrial civilization, humanity has continuously enhanced its capabilities by developing and innovating new products, promoting the optimization and upgrading of production systems to align with the advancement of productive forces. From the Stone Age through the Bronze Age to the Iron Age, product innovation was heavily dependent on the application of new materials (material revolution). During the First Industrial Revolution, characterized by the steam engine, and the Second Industrial Revolution, marked by the electric motor, the focus of product innovation transitioned from materials to power and energy structures (power revolution). By replacing repetitive manual labor with new types of power machines, industrialization and automation emerged as the primary drivers of social production development. With the substantial enhancement in productivity, the volume of information managed by humans has expanded exponentially. Consequently, the challenge of efficient information processing has emerged as the primary obstacle hindering further productivity advancements. There is now an urgent imperative to develop innovative technologies and products that can facilitate the automation of certain cognitive tasks, transcend the limitations of human intelligence, and redirect intellectual resources towards more valuable and creative endeavors (intellectual revolution)[16-17]. In conclusion, from the material revolution to the power revolution and on to the intelligence revolution, intelligence has emerged as the future direction of human product innovation and development.

From the perspective of technological philosophy, the essence of the aforementioned product innovation and evolution process is rooted in the advancement of the manufacturing system. Initially, the manufacturing system was characterized by a "Human-Physical" system (HPS) where humans and physical products were co-constructed. With the advent of information technology, this evolved into digital manufacturing within a "Human-Cyber-Physical" system (HCPS1.0). Further propelled by advancements in new-generation artificial intelligence, the system is now transitioning towards next-generation intelligent manufacturing (HCPS2.0), which incorporates advanced intelligent features such as self-perception, self-adaptation, and self-decision-making[17]. Through the stages of holistic manufacturing systems, it can be observed that product innovation evolution has the following characteristics: ①Passive to active, in the HPS system, product innovation is more dependent on optimizing and improving existing products. The product operation is highly obedient to human instructions and lacks autonomy. In the HCPS system, product innovation is more dependent on the understanding and reuse of principle knowledge, forming original and innovative solutions to specific problems, improving the intelligence level of products, and integrating multi-factor comprehensive decision-making while accepting human instructions, thereby possessing certain autonomy. ②Discrete to integrated, in the HPS system, product innovation involves relatively independent physical processes, and the structural modules of products are strictly corresponding to specific functional requirements. In the HCPS system, product innovation involves numerous nonlinear and time-varying physical processes, forming a complex mechatronic system with rich functions, highly coupled, and dynamically open. ③Disembodied to embodied, in the HPS system, the relationship between people and products tends to be binary and opposed. People create products and control and use them through various means, which is essentially a discrete, linear, highly separated human-machine relationship. In the HCPS system, the relationship between people and products tends to be collaborative. Intelligent products assist humans in sensing, learning, and decision-making, and a high-level human-machine collaborative intelligent manufacturing system is established.

## 2.2 The Development of New-Quality Productive Forces Calls for Iconic Product Innovation

New-quality productive forces are those stimulated by breakthroughs in disruptive technological innovations in key fields. Essentially, technological innovation is regarded as the driving force for the development of social production, promoting the upgrading of productive forces from "quantitative change" to "qualitative change" levels[18]. Disruptive technological innovation refers to groundbreaking and transformative technological advancements that surpass existing market-leading technologies in multiple dimensions, including principles, efficiency, and cost. Such innovations can trigger industrial revolutions and transform production methods. In contrast, sustaining technological innovation involves incremental improvements and optimizations of existing technologies along their original performance trajectories through localized process enhancements[19]. The product represents a hierarchical and intricate integration of multi-functional and cross-disciplinary technological innovations, which significantly influences the technological advancement and industrial competitiveness in relevant fields. Consequently, to cultivate high quality productive forces, it is imperative to elucidate the intrinsic philosophical mechanisms underpinning both traditional and iconic products and to scientifically formulate research and development strategies. Drawing on systems science and the philosophy of technology, this paper provides an in-depth analysis.

Traditional products are extensively utilized in sectors such as mechanical manufacturing, aerospace, and transportation. By logically transforming and analyzing engineering problems, design constraints and boundary limitations are clearly defined, production experience and disciplinary knowledge are integrated, the functional structure of the product is accurately mapped, and quality and reliability are ensured comprehensively. The innovation of traditional products is grounded in the HPS system, where products serve as tools for humans to solve problems. Essentially, this process involves the reuse and creation of knowledge for specific challenges. However, there are notable drawbacks in this process: ① Static Perspective. Engineering problems are often defined through static observation and description, relying on static knowledge that has been encapsulated and stored. This approach overlooks dynamic changes in the real world, leading to a lack of self-organization and adaptive capabilities in products when dealing with complex environments; ② Incomplete Knowledge Base. Traditional product innovation heavily depends on explicit knowledge that can be represented, while tacit knowledge—such as intuition, habits, and intentions—is either ignored or abandoned due to its unrepresentability. This results in an incomplete knowledge base for product innovation. Moreover, traditional product innovation adheres to Cartesian dualism, emphasizing formal logical thinking and symbolic intelligence based on computational representation, which leads to limited autonomy in products.

Iconic products, serving as the primary drivers for the advancement of future industries, represent an integration of novel theories, cutting-edge technologies, and innovative methodologies. Leveraging advanced and interdisciplinary knowledge, they foster a wealth of original achievements and disruptive technological innovations, thereby ensuring the establishment of a robust future industrial system architecture and the cultivation of high-quality productive forces. The innovation of iconic products is based on the HCPS system. Products are regarded as "cooperative objects" and "universal partners" of human beings[20]. The essence is to create "human-like" intelligent agents and endow products with self-organizing capabilities similar to the body (embodiment). Compared with traditional products, iconic products possess several distinct advantages: ①Dynamic perspective. Iconic products represent a highly integrated and innovative fusion of advanced technologies and hierarchical systems, resulting in superior integration effects and emergent functionalities. This enhances the autonomy, flexibility, and adaptability of products, enabling them to efficiently respond to complex and dynamic external environments. ② Comprehensive knowledge management. Grounded in human-machine integrated intelligent systems, iconic products excel in capturing, transferring, transforming, and managing tacit knowledge dynamically. This strengthens the product's ability to autonomously learn and generate knowledge, achieving self-optimization and self-improvement in knowledge management. Consequently, they establish a comprehensive knowledge system that integrates both explicit and tacit knowledge. Adhering to the phenomenological philosophy of mind-body unity, Iconic products balance the precision and efficiency of symbolic logic computing with the integration and comprehensiveness of system self-organization, advancing towards autonomous and embodied intelligence characterized by strong autonomy and synergy[21].

In conclusion, the innovation of iconic products serves as both the focal point and the breakthrough for the development of new quality productive forces. By adhering to a phenomenological systems thinking that integrates mind and body, we can address external challenges from a dynamic perspective and through a comprehensive knowledge framework. This approach fosters the development of highly autonomous embodied intelligence, spurs numerous disruptive technological innovations, and ultimately drives the high-quality advancement of new quality productive forces.

## 3 The First Principle of Knowledge Phenomenology in Iconic Products Intelligent Innovation

The intelligent innovation of iconic products is a creative activity grounded in comprehensive knowledge. By leveraging this comprehensive knowledge, new principles, methods, and technologies are generated to facilitate the realization of conceptual designs and achieve original innovation. To understand the common patterns of intelligent innovation in iconic products, it is crucial to conduct a fine-grained analysis of the underlying logic and system framework of the product innovation process, as well as to analyze and summarize the scientific evolution processes and fundamental principles that drive such innovations. Therefore, this paper focuses on "intelligent innovation" as the central theme and employs the phenomenological method of essential reduction to deeply examine the internal logical framework of iconic product innovation, thereby revealing the philosophical principles governing the generation and evolution of comprehensive knowledge.

## 3.1 Comprehensive Knowledge Support for Iconic Product Intelligent Innovation

The intelligent innovation of iconic products is grounded in the HCPS system. Leveraging advanced technologies such as artificial intelligence, cloud computing, big data, nanotechnology, and Internet of Things, it integrates data sensing, human-computer interaction, decision-making control, and adaptive feedback. This integration enables comprehensive collaboration and intelligence across the entire product lifecycle, from R&D design through production, manufacturing, transportation services, and sales management. It optimizes material, information, and energy flows within smart manufacturing systems, ultimately creating advanced intelligent products for the modern manufacturing IoT network. ①At the technical level, intelligent innovation is the integrated fusion of intelligent technology and manufacturing technology. Through intelligent technology, the ability of product information perception, collection and control is enhanced, and through manufacturing technology, the physical performance parameters of products are optimized to implement high-quality product innovation projects. ②At the system level, intelligent innovation integrates the social system into the manufacturing system, establishing a complex social-technical system that integrates multiple key elements such as humans, machines and the environment. It utilizes multidisciplinary knowledge such as philosophy, mathematics, computer science, mechanical engineering, and management to cross-fuse physical systems, information systems and social systems, promoting products to move from the original closed innovation to open innovation and embedded innovation. ③At the philosophical level, intelligent innovation focuses on the underlying structure of human intelligent behavior, discovers that symbolic intelligence and logical intelligence are based on the emergence of physical perception and experience, and establishes embodied intelligence with loose coupling, distribution and strong autonomy, constantly approaching and aligning with humans.

Intelligence, in its connotation, represents the integration of knowledge and cognitive processes, while in its extension, it denotes the capability to discover and apply principles, analyze and solve problems[17]. The core of intelligent innovation lies in leveraging artificial intelligence (AI) technology to transfer these capabilities to products, thereby endowing them with human-like intelligent features and alleviating the intellectual burden on humans. From a historical perspective, early AI development was rooted in symbolism, which advocated for the creation of physical symbol systems to simulate human brain functions. This approach relied on rule-based reasoning derived from explicit knowledge. For instance, classic expert systems were often limited to providing explicit solutions for pre-defined problems within a narrow scope, exhibiting weak generalization capabilities. In contrast, contemporary AI primarily stems from connectionism, which emphasizes mimicking the biological structure of the human brain by constructing large-scale parallel neural networks to emulate human intelligence. Additionally, some scholars, inspired by behaviorism, achieve autonomous control and evolution of intelligent agents through direct interaction with the external environment and real-time updating and iteration of internal information[22].

Knowledge serves as the foundation and prerequisite for intellectualization. The acquisition and application of knowledge permeate the entire process of product innovation, embodying the materialization of human wisdom in manufacturing activities[17]. Polanyi emphasized that tacit knowledge provides the backdrop and precondition for acquiring explicit knowledge[23]. The intuitive and emergent characteristics inherent in tacit knowledge form the basis for constructing cognition and supply the raw materials for information acquisition and logical deduction. The Limitations of Symbolism lies in its disregard for the fundamental structure of human knowledge generation and application, confining itself entirely to the logical framework of computational representation of explicit knowledge. Conversely, connectionism and behaviorism excel by capturing certain aspects of tacit knowledge and generating new knowledge based on it, exemplified by the emergence mechanisms in neural networks and the external interaction behaviors of intelligent agents. Therefore, the crux of intelligent innovation in iconic products hinges on accurately targeting the ontological structure of human intelligence, establishing a comprehensive knowledge engineering system that integrates both explicit and tacit knowledge, endowing products with capabilities for autonomous knowledge generation, reasoning, and reuse, thereby facilitating the transition from experience-based design to knowledge-driven innovation.

### 3.2 The First Principle of knowledge Phenomenology

The philosopher Russell held that "knowledge is not a precise concept [24]." In academic discourse, there are extensive discussions and definitions of knowledge: Plato posited that knowledge is justified true belief; Marx argued that knowledge is a product of social practice and is validated through such practice; from an economic perspective, knowledge is considered a valuable outcome of human labor and serves as a production factor; from the viewpoint of information theory, knowledge represents the accumulation and structuring of information for specific purposes[25]. Although the definitions of knowledge are diverse and complex, lacking a universally accepted consensus, the essence of knowledge can still be understood through several common characteristics: ① Knowledge represents the cognitive synthesis that humans derive from practice regarding the phenomena and fundamental laws of nature, society, and even thought; ②Knowledge is a significant intellectual achievement of humans and possesses a degree of objectivity; ③ Knowledge is generated, transmitted, and reused in dynamic processes, while also manifesting as static, structured knowledge products[26]. In conclusion, the concept of knowledge is further refined at the philosophical level: knowledge represents the crystallization of cognition, encompassing both the processes and outcomes of human understanding of both the spiritual and material worlds.

Phenomenology is a theoretical method for essential reduction of subjective experience and objective phenomena. By suspending preconceived empirical judgments, it scientifically describes the substantive content of the research object. In this paper, combined with the theoretical method of spatiotemporal phenomenology[19], integrating the original spatiotemporal dimension and the characteristics of human practical behavior, a systematic analysis of the structure of knowledge generation and evolution is conducted, and the concept of knowledge phenomenology is established.

The phenomenological giant Edmund Husserl, starting from the temporal dimension of human experience and the structure of consciousness, proposed the structure of "internal time-consciousness" comprising "urimpression, retention and protention." The "urimpression" refers to the recognition and analysis of the perceptual material of the present moment; "retention" refers to the continuity and grasping of past perceptual content; "protention" refers to the anticipation and intentionality toward future events. The internal time-consciousness structure forms the foundation of human perceptual experience, where each "now" time-point encompasses partial perceptual content and intentionality of both the "past" and the "future." These three elements are highly integrated, presenting a triadic cognitive structure of "past-present-future." Specifically, the cognition of objects is grounded in the retention of past events and the protention of future events, integrating the perceptual materials of the current primordial impression to form an overall cognitive framework. Objective time divides events into discrete moments, creating fundamental heterogeneity between these moments, which necessitates active recall and prediction, thereby significantly increasing the cognitive load on the subject and disrupting the continuity of the cognitive process and the coherence of perception[27-28]. The literature indicates that tacit knowledge originates from perceptual experiences and sensations acquired through bodily movements and sensory stimuli in practical life[26]. Consequently, inner temporal consciousness, as a prerequisite for the formation of perceptual experiences, underpins the generation and evolution of tacit knowledge and serves as the foundation for knowledge acquisition, application, and creation. This stage represents the initial composition of human knowledge, primarily comprising empirical and experiential tacit knowledge, supplemented by explicit knowledge derived from describable and representable perceptual experiences to form preliminary cognition.

Furthermore, the aforementioned tacit knowledge evolves over time, continuously transforming and developing into a complex knowledge structure. Dreyfus's analysis of "expertise knowledge" acquired by experts illustrates this process in a structured and systematic manner. The acquisition of expertise knowledge is divided into seven stages: novice, advanced beginner, competent, proficient, expert, master, and practical wisdom[29]. The initial three stages focus on learning and applying established rules and content, where individuals acquire explicit knowledge based on their existing tacit knowledge, combined with abstract and conceptual logical reasoning. In contrast, the latter four stages emphasize embodied and contextualized repetitive practice, which facilitates the transformation of explicit knowledge into tacit knowledge and the development of pre-reflective and pre-conscious physical skills (expertise knowledge), such as skilled workers controlling part processing temperatures and experienced pilots performing piloting tasks[30-31]. Thus, the knowledge evolution process can be summarized as follows: ① Knowledge generation, where human beings form initial perceptual experiences and acquire tacit knowledge through intrinsic temporal consciousness; ②Tacit to explicit conversion, wherein some tacit knowledge is transformed into explicit knowledge via rational logical thinking and symbolic representation; ③ Explicit to tacit internalization, where explicit knowledge is re-internalized into new tacit knowledge through physical practice. At this stage, the newly generated tacit knowledge is not entirely distinct from explicit knowledge but represents an embodied transformation of the original knowledge; ④ Integration of explicit and tacit knowledge, which involves the cross-temporal and cross-scenario application of both types of knowledge, promoting their organic integration. This process establishes an efficient and high-quality comprehensive knowledge system, leading to expertise characterized by practical skills and adaptability to complex environments.

From an engineering science perspective, a significant limitation of current artificial intelligence technology is its reliance on symbolic operations to mimic expert decision-making processes. It fails to transform explicit knowledge into tacit knowledge through physical practice, thereby preventing the acquisition of true "expertise knowledge"[32]. Consequently, this paper proposes the establishment of an "embodied expertise knowledge system" grounded in inner time consciousness. ①Establish a knowledge acquisition mechanism aligned with intrinsic temporal consciousness. The concepts of "protention" and "retention" embody the cognitive intentionality and capability to perceive objective entities across temporal domains[27]. By acquiring multi-dimensional and multi-scenario cognitive data, a trinity of foundational cognitive structures is formed. Through the development of high-precision intelligent agent models adaptable to multiple objectives and variables, the efficiency of cross-temporal domain knowledge acquisition is enhanced, thereby promoting integrated demand perception, agile response, and information integration, while reinforcing the knowledge generation and autonomous evolution capabilities of intelligent systems. ② Facilitate the explicitation of tacit knowledge leveraging big data and artificial intelligence, fully utilizing advanced information technology to accelerate the processing of multi-source heterogeneous information, perform cognitive computing for thought and feature extraction, and conduct research on knowledge representation and reasoning using methods such as ontology models, knowledge graphs, and semantic networks. ③ Facilitate the tacitization of explicit knowledge through physical practice. By integrating intelligent agents into mobile terminals of smart equipment such as CNC machine tools, humanoid robots, and autonomous vehicles, a large-scale, distributed, and parallel multi-agent system (MAS) is established. This MAS supports knowledge transfer and human-machine collaborative operations in heterogeneous network environments, enabling real-time perception of the physical environment and acquisition of embodied and contextualized tacit knowledge. ④ Develop an intelligent decision-making method driven by a hybrid of data and knowledge. Through iterative practice to accumulate empirical data, the knowledge base and intelligent decision-making system are continuously updated and refined. This process enhances the system's autonomous decision-making and proactive behavior capabilities, ultimately acquiring expertise characterized by specialized skills.

In conclusion, the knowledge Phenomenology integrates the original spatiotemporal dimension and bodily practice, offering a systematic philosophical framework for understanding the processes of knowledge generation and evolution. Driven by inner temporal consciousness and grounded in multi-spatiotemporal perceptual experiences, an initial cognition primarily composed of tacit knowledge and supplemented by explicit knowledge is formed. In response to the demands of production and life practices, this cognition undergoes a dynamic transformation where tacit knowledge becomes explicit and explicit knowledge becomes tacit. This process integrates logical thinking with bodily perception, promoting a high-level synthesis of explicit and tacit knowledge, ultimately leading to expertise based on embodied skills (Figure 1).

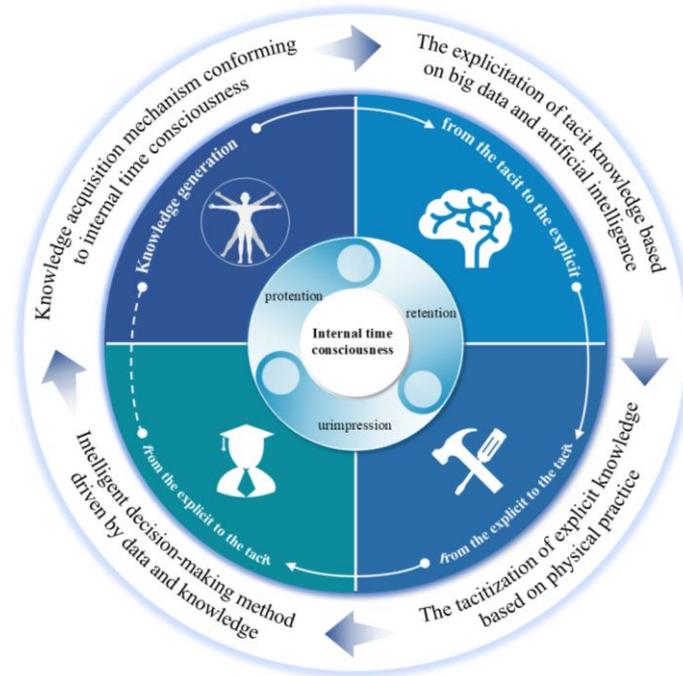

Figure 1 The First Principle of Knowledge Phenomenology

## 4 Knowledge Phenomenology Driven Forward Innovative Design of Iconic Products

Forward design relies on precise and self-consistent scientific theoretical guidance, integrating critical processes of product design such as demand definition, functional decomposition, and system synthesis to achieve comprehensive independent innovation from concept to product. By combining innovative thinking, this approach integrates principle, method, technology, and service innovations into a cohesive framework, enabling forward innovative design for future equipment products and facilitating the transition of China's foundational equipment products from imitation to creation and leadership [33-34]. Iconic products represent the pinnacle of value in international scientific and technological competition and are central to the national economy. To systematically enhance the core competitiveness of these products, it is essential to start from their inherent philosophical mechanisms and scientific principles. Therefore, based on the first principles of knowledge phenomenology, this paper develops a forward innovative design methodology for iconic products, providing theoretical support for achieving disruptive innovation.

### 4.1 Phenomenological Reduction of the Evolutionary Principles in Product Design

From the perspective of the development process of human civilization, the evolution of design can be divided into the following three stages. ①Traditional design in the era of agricultural civilization (Design 1.0), whose essence is intuitive design and inspiration design based on the HPS system. It uses simple tools and manual methods, and focuses on the optimization of product appearance and shape. Limited by the development level of productive forces, only human and animal power, as well as small-scale wind and water power can be utilized. The accumulated design experience and theoretical knowledge are very limited, and no breakthroughs can be made in efficient and precise product design and manufacturing. ②Modern design in the era of industrial civilization (Design 2.0), in essence, is empirical design and integrated design based on the HCPS 1.0 system, and has developed design technologies and methods such as digital modeling, virtual simulation, comparative experiments, and CAx. The mechanization and industrialized large-scale production led by the Industrial Revolution provided a solid material and information foundation, extensive market demand traction, and rich theoretical and practical experience for design. By summarizing common empirical laws and integrating multidisciplinary knowledge and principles such as engineering and management, a number of scientific and engineering modern design theoretical methods such as the TRIZ theory of inventive problem-solving, QFD design, and axiomatic design emerged. ③The innovative design in the era of knowledge network (Design 3.0), in essence, is the original and disruptive design based on the HCPS 2.0 system, using technical methods such as artificial intelligence, virtual reality, and digital twin for knowledge innovation and principle innovation. The information revolution and intelligence revolution have made human beings' ability to master knowledge unprecedentedly powerful. Digitalization, networking, and intelligence have become important characteristics of social and economic development, providing rich intellectual resources for innovative design, continuously stimulating human creativity, and designing iconic products with huge market potential[35-36].

By reviewing the above process and considering the current state of product design development in China, the following trends can be summarized. ① From perceptual intuition to rational synthesis. In the traditional design stage, due to the lack of support from modern science and engineering theories, the design mainly relied on human instinctive intuition and inspiration; in the modern design stage, the proportion of experience in the design process increased significantly, and it was necessary to strengthen induction and summary based on rational thinking, but still lacked in-depth discussion of design principles; in the innovative design stage, higher requirements were put forward for principle innovation, and it was necessary to fully apply rational logic, master the scientific essence behind empirical rules, and carry out disruptive design. It should be noted that this trend is a manifestation of the increasingly scientific and standardized human design activities and innovative thinking. Rational synthesis does not abandon perceptual intuition, but fully realizes the importance of the latter, and studies and applies it in combination with scientific methods, integrating the dual advantages of rationality and perception. ② From reverse reasoning to forward innovation. In the modern design stage, there are a large number of reverse designs mainly based on tracking and imitation. Through digital processing of advanced products, the structural characteristic parameters of the original products are analogically reasoned, and small-scale optimization and improvement are implemented; in the innovative design stage, it is necessary to start from the fundamental scientific principles, realize the physical transformation from high-level design intentions to physical products, precisely control product performance, and create high value-added products. ③ From single serial to system parallel. In the early stages of traditional design and modern design, product functional requirements and application scenarios are relatively simple. By decoupling design problems according to the single-discipline serial research and development method of "dividing into parts", several sub-tasks are defined, and simple structural permutations and combinations are used to implement technology integration; in the middle and later stages of modern design and the innovative design stage, products evolve into complex electromechanical systems that adapt to multi-functional requirements and scenarios. It is necessary to carry out top-down research from the perspective of system science and engineering, grasp the highly complex coupling relationships among subsystems, structures, and processes, and use advanced information technology and cross-time and space multi-dimensional information to implement concurrent engineering of product design and reveal and represent the potential knowledge laws within the products[20].

Design is a creative intellectual activity through which humans transform nature, and its essence lies in the process of knowledge flow, integration, competition, and evolution[37]. By applying phenomenological reduction to knowledge, we can deeply identify the underlying philosophical mechanisms driving the evolution of product design. ①The transformation from perceptual intuition to rational synthesis essentially represents the conversion of knowledge from implicit to explicit. Traditional design grasps design problems and demands in life experiences through perceptual intuition, essentially obtaining the tacit knowledge derived from production and life practices. Modern design and innovative design, on this basis, employ rational logical thinking to transform vague and uncertain design problems and descriptions into clear and definite scientific problems and objective constraints, precisely mapping the corresponding design knowledge and promoting the explicitization of tacit knowledge. ②The transformation from reverse reasoning to forward innovation essentially represents the conversion of knowledge from explicit to implicit. Reverse reasoning is to extract features from existing products, essentially mining the corresponding explicit knowledge through the characterization of product parameters. Forward innovation still incorporates reverse thinking[20]. On the basis of extracting explicit knowledge, it deeply recognizes the underlying tacit knowledge of principles and promotes the implicitization of explicit knowledge through continuous design practices. ③The transformation from single serial to system parallel is essentially the cross-temporal and explicit-implicit integration. The single serial is based on the static, causal and discrete computational representation thinking, showing a simple linear spatiotemporal relationship of input/output, with sparse knowledge distribution characteristics and a low level of association between explicit and implicit knowledge. The system parallel adheres to the dynamic, systematic and continuous phenomenological thinking, showing a non-timely response cross-temporal cascade, integrating knowledge discovery, fusion and creation, and promoting the cross-temporal integration of explicit and implicit knowledge[38].

**4.2 Forward Innovative Design of Iconic Products**

Based on the aforementioned analysis, the evolution trend of product design is closely aligned with the process of knowledge evolution, which can be interpreted as the manifestation of knowledge phenomenology in design practice. Therefore, to achieve forward-looking innovative design of products, it is essential to thoroughly understand the philosophical underpinnings of knowledge phenomenology and stimulate disruptive innovation of iconic products from the foundational level of knowledge. Furthermore, this paper integrates the first principle of knowledge phenomenology with the design practice process to construct a theoretical framework for forward-looking innovative design of iconic products. This framework is based on the step-by-step mapping and resolution of "design problem space - explicit knowledge space - tacit knowledge space - innovative solution space."

4.2.1 Multi-source Demand Coupling Driven Design Problem Space Construction

Problem definition marks the inception of the design activity and is fundamentally an open, iterative process that integrates multiple demands, knowledge, and experience[39-40]. Traditional product design narrowly focuses on market demands to define the design problem. It gathers extensive user data, establishes a user tagging system and structured analytical models, accurately identifies user characteristics, delineates design constraints, and creates a design and manufacturing framework centered on user needs with rapid market responsiveness. Different from traditional products, iconic products are the key to enhancing the national scientific and technological innovation level and industrial competitiveness, and need to be developed from the following aspects. ①At the macro level, the problem definition of innovative design of iconic products should be guided by national strategic deployments such as innovation-driven development and advanced manufacturing, facing the demands of cutting-edge technological competition, national economic development and national defense construction, and improving the informatization, intelligentization and domestic production capacity of products. ②At the meso level, it is still necessary to actively respond to dynamic market demands, grasp the diversified and individualized development trends of user demands, and use technologies such as ontology modeling, association modeling and psychological cognition to deeply explore users' latent demands and product expectations. ③At the micro level, it is necessary to combine the actual industrial production conditions, use technologies such as the Internet of

Things, the Internet of Knowledge and the Internet of Body, and mine high-value information from a large amount of structured, semi-structured and unstructured data, and integrate technical conditions and resource allocation to granularize the macro and meso problems into design constraints. Thus, based on the "macro-meso-micro" perspective, a design problem space of multi-source demand coupling of "national strategy - market demand - industrial production" is constructed, and a forward innovative design model driven by actual major engineering demands is established.

From the perspective of knowledge phenomenology, the establishment of the problem space in design represents a critical stage in knowledge evolution where new knowledge is generated. The delineation of this problem space encompasses physical, information, and social systems, integrating vast and intricate human-machine-object interaction processes. These interactions, transmissions, and stimulations result in deep-level coupling and integration, as well as information cascades, leading to the emergence of a significant amount of tacit knowledge. On one hand, the intuitive and non-computational nature of tacit knowledge introduces significant uncertainty and ambiguity into the design problem space. On the other hand, tacit knowledge integrates multi-system and multi-level requirements, providing a more comprehensive and nuanced depiction of interrelationships between entities, thereby enhancing the positioning and prediction of future market trends. Furthermore, leveraging the internal time consciousness mechanism inherent in human tacit knowledge generation, large-scale and high-precision intelligent agent models capture multi-system, multi-field, and strongly nonlinear cross-temporal information. This process refines the multi-source requirement integration cognitive architecture, continuously optimizing the design problem space and offering forward-looking guidance for innovative design.

4.2.2 Big Data Intelligence Driven Explicit Knowledge Space Construction

How to utilize big data and artificial intelligence technologies to sort data and eliminate noise from the complex and disordered discrete information units, and establish an ordered knowledge system for specific design problems is the primary step in solving design problems. Specifically, the sources of design knowledge at this stage can be divided into the following three aspects: ① Ready-made related equipment products, which are essentially integrated solutions and scientific representation structures for specific design problems, contain large-scale and high-quality advanced technical theories and methods, and are a concentrated reflection of long-term engineering practice wisdom. Through product resource retrieval based on big data, a product database for basic design tasks is established, providing a preferred choice for design knowledge reuse. ② Production event log information, which is essentially semi-structured data resources based on industrial production business processes and is the basis for establishing control flow and process models[41]. Through the composite event processing technology (CEP) based on the event-driven architecture (EDA), the system operation status and data flow are monitored in real time, and a large amount of knowledge and experience based on actual production practices are mined. ③ Social network data, which is essentially multimodal data resources based on human social systems and daily life monitoring, is a structured mapping of real physical systems, social systems and virtual information systems. Through the Internet of Things and edge computing technologies, multimodal data information such as images, voices and videos is widely collected, providing a solid data foundation for design knowledge mining.

From the perspective of knowledge phenomenology, the establishment of explicit knowledge space represents a critical phase in the evolution of knowledge from implicit to explicit. On one hand, by leveraging big data and multi-source heterogeneous information mining, this process facilitates the integration of generalized design knowledge across multiple fields and disciplines. It tracks uncertain factors in the quantitative design process, thereby providing a robust knowledge foundation for lean and intelligent design in complex service environments. On the other hand, a significant amount of tacit knowledge within the design problem space is articulated through data representation and comparative analysis. The inherent ambiguity of design problems is mitigated and simplified via data-driven approaches, resulting in precise and actionable design constraints and criteria, which significantly enhance the scientific rigor and logical coherence of the design process. The mapping from the design problem space to the explicit knowledge space relies on a thorough understanding and analysis of the explicit knowledge set. Through intelligent big data representation and computational identification, the primary functional requirements, management models, and usage scenarios of the product are clearly defined, enabling an organized, standardized, and modularized approach to the design process.

4.2.3 Embodied Expertise Driven Tacit Knowledge Space Construction

The core of forward-looking innovative design for iconic products lies in driving the transformation and upgrading of the product's intrinsic knowledge structure and establishing a comprehensive knowledge system that deeply integrates explicit and tacit knowledge. The sources of tacit knowledge can be categorized into two main aspects: ①Cognitive tacit knowledge, which forms the fundamental cognitive framework of human beings, reflecting their perspectives and attitudes toward the external world, typically expressed as concepts, intentions, and expectations; ②Skill-based tacit knowledge, which stems from long-term practical experience and represents the integration of mental awareness and physical memory, usually manifested as skills and techniques[25]. In the process of product design, acquiring cognitive tacit knowledge is crucial for constructing the design problem space. This involves developing an objective understanding and accurate representation of external design requirements and challenges. On the other hand, obtaining skill-based tacit knowledge is essential for building the tacit knowledge space. By leveraging human-robot skill transfer technology (HRST), certain physical skills can be imparted to equipment products, enabling them to acquire some tacit knowledge and human-like operational capabilities. Specifically, HRST technology is a key component of the HCPS system, encompassing three general processes: teaching, model learning, and task reproduction. Additionally, there are three primary methods of skill transfer: vision-based, teleoperation, and physical interaction[42]. Based on HRST technology, certain human physical skills are systematically generalized and formally described, and subsequently transferred to relevant products via kinematic programming. The integration of human advantages such as adaptability, intelligence, and dexterity with machine attributes like high efficiency, high precision, and high compliance significantly enhances the products' capabilities in learning, storing, and applying implicit knowledge.

From the perspective of knowledge phenomenology, the establishment of the tacit knowledge space represents a critical phase in the evolution of knowledge from explicit to tacit. The construction of the explicit knowledge space aims to transform ambiguous and divergent design problems into clearly defined design constraints, systematically delineating the primary functions and application scenarios of the product while ensuring the feasibility of engineering design. However, this process inevitably results in the loss of some potential design requirements. Additionally, there are tendencies towards over-simplification, discretization, and knowledge obsolescence in the design definitions. The construction of the tacit knowledge space, facilitated by HRST technology, enables the transfer of certain tacit skills to the product through methods such as dynamic modeling and motion programming. Essentially, it leverages existing explicit knowledge to emulate and approximate human physical skills, thereby expanding product functionality based on embodied expertise.

4.2.4 Human-Machine Collaborative Decision-Making Driven Innovation Solution Space Generation

Human-machine collaborative decision-making is inevitable for the innovative design of iconic products. The specific reasons can be divided into the following two aspects. ①In the HCPS 2.0 era, the human production mode has undergone tremendous changes, which have a significant impact on the generation and reuse of expertise knowledge. Under the traditional production mode, professional and technical talents constantly discover and solve problems based on production tasks, combined with expertise knowledge and work experience. In the context of HCPS 2.0, professional and technical talents no longer directly face the traditional production environment dominated by physical manufacturing equipment, but rather a large volume and wide variety of process data[43]. The original expertise knowledge based on the physical environment cannot directly play a role. It is necessary to extract objective phenomena and data characteristics suitable for expertise knowledge judgment through large-scale data analysis by computers to form a reasonable solution based on human-machine collaborative decision-making. ② Human practical activities are extremely rich, have infinite development, and the relationships between people are extremely complex. There is no possibility of developing a purely autonomous unmanned creation system[44]. To promote the innovative design of iconic products, it is necessary to give full play to the advantages of human creative thinking, integrate theoretical thinking, image thinking and inspiration thinking, utilize innovative means and methods based on human-machine combination and human-oriented, promote the high-quality integration of explicit/implicit knowledge, enhance human thinking ability and innovative design ability, and make it possible for products to achieve disruptive, breakthrough and fundamental innovations[45].

Furthermore, the development of an innovative solution space through human-machine collaborative decision-making can be advanced both vertically in theoretical frameworks and horizontally in technical methodologies. ① At the vertical theoretical level, adhering to Qian Xuesen's "Comprehensive Integration Method from Qualitative to Quantitative for Open Complex Giant Systems": First, define the product design tasks based on the problem space; Second, integrate explicit and tacit knowledge spaces by incorporating expert judgments to establish objective cognition; Third, construct a design problem-solving system grounded in knowledge engineering that synthesizes the insights and solutions of designers, engineers, and experts, generating innovative solutions through iterative refinement based on both subjective and objective evaluations[44]. ② At the level of horizontal technical methods, the development of technologies from hybrid human-machine collaboration to multi-person and multi-machine collaboration aims to fully leverage the collective wisdom of human groups in the design process. Hybrid human-machine collaboration dynamically optimizes decision-making by creating a closed loop that tightly integrates humans, multiple intelligent terminals, and smart machines. This enables continuous feedback on human actions and behaviors, real-time training and optimization of intelligent algorithms, and establishes a "human-in-the-loop" machine learning framework, thereby significantly enhancing the intelligence of design decisions. Multi-person and multi-machine collaboration expands the scope of shared resources supporting decision-making across spatial dimensions. By establishing a distributed multi-agent system that integrates software, hardware devices, and professionals, it overcomes the limitations of traditional single-expert AI systems. This approach facilitates multi-agent collaborative and parallel problem-solving for design challenges, pooling collective wisdom and fostering collective innovation.

From the perspective of knowledge phenomenology, the essence of generating the innovation solution space is the cross-temporal and explicit-implicit integration. Human-machine collaborative decision-making is an organic integration of human-machine-objects and numerous complex physical processes, integrating the explicit/implicit knowledge space to form an open complex giant system with advanced intelligent features and innovation levels. This process shows significant cross-temporal characteristics. On the one hand, the essence of the integration of explicit and implicit knowledge is the cross-temporal comprehensive human cognition. By integrating theoretical knowledge and practical experience from multiple sources of temporal and spatial contexts, it provides a flexible, dynamic and large-scale knowledge base for decision-making. On the other hand, the internal of the complex system is not the minimalist feedback and linear causal relationship in the traditional cybernetics. It breaks through the continuous factual logical calculation within the traditional single temporal and spatial interval and forms a deep cognition based on multiple levels and multi-dimensional temporal and spatial.

Retracing the forward innovative design process of iconic products, it becomes evident that this process has thoroughly undergone knowledge generation grounded in phenomenology, the explicitation of tacit knowledge, the tacitization of explicit knowledge, and the integration of both forms of knowledge. This has established a comprehensive knowledge system tailored to specific production and life tasks, thereby positively stimulating independent knowledge innovation in iconic products from a philosophical perspective.

# 5 The Reshaping of the Global Future Industry Competitive Landscape through Iconic Product Innovation

The essence of forward innovative design of iconic products lies in the engineering application of knowledge phenomenology. Grounded in the universal principles of human knowledge evolution, a smart and lean product design framework is established, progressing from macro to micro and from theory to practice. The successful implementation of advanced technology is an indispensable prerequisite for this engineering application. By decoupling the innovation solution space in reverse and integrating the principles of knowledge phenomenology, we can systematically predict the disruptive technologies required for iconic product innovation, moving from micro to macro. This approach, which combines forward innovative design with reverse decoupling prediction, establishes an innovative system engineering framework for iconic products. This framework supports the scientific analysis of global future industrial competition and proposes embodied intelligence enhancement strategies for iconic products.

## 5.1 Prediction of Disruptive Technological Innovation Based on the Reverse Decoupling of Iconic Products

According to the "Function-Behavior-Structure" (FBS) design theoretical model, the design process can be characterized by the specific mapping relationships among product functions (F), behaviors (B), and structures (S). Consequently, utilizing the FBS design model enables precise description of the common characteristics of the design object, facilitating the establishment of a design knowledge representation method based on ontology[46]. This paper conducts reverse hierarchical decoupling based on the FBS design ontology for the innovation solution space. Combined with the first principle of knowledge phenomenology, it explores the knowledge basis and spatiotemporal application scenarios required for technological innovation and credibly predicts future disruptive technological innovations in industries. Further, by integrating the forward innovative design of iconic products and the reverse decoupling prediction process of disruptive technological innovation, and in combination with the "V model" of systems engineering, integrating "top-down system concept requirement decomposition" and "bottom-up system integration and verification"[47], the innovative systems engineering of iconic products is established (Figure 2).

### 5.1.1 Construction of the Technological Problem Space Based on Spatiotemporal Migration

Time and space are the most important and fundamental attributes of the real world. All data, knowledge and information can be characterized from the two dimensions of time and space[48]. From the perspective of systems science, the innovation solution space is the knowledge fusion and creation for specific design problems, requirements and tasks, and it is a complex knowledge system including multiple levels of time and space dimensions.

In the prediction phase of disruptive technological innovation, the emphasis is on driving the implementation of innovative design solutions through technological research. It is crucial to integrate specific engineering practice requirements, facilitate the reorganization of foundational knowledge and data, and adapt to complex working conditions and new technology service environments. The core of this process lies in mapping the "innovation solution space" to the "technical problem space" via transformations in spatiotemporal representation methods, thereby establishing a "design-engineering" task transformation framework based on spatiotemporal migration. Specifically, it can be divided into the following two aspects: ① Transformation of the knowledge spatiotemporal representation architecture. By establishing a high-precision and high-resolution global spatiotemporal information system architecture for complex manufacturing environments, and taking the spatiotemporal representation method of the industrial production environment as the benchmark, the knowledge expressions of the original strategic information, market information and production information are unified, and a product modeling method integrating spatiotemporal coordinates and object motion characteristics is developed. ② New knowledge acquisition based on the new spatiotemporal representation architecture. The analysis of specific industrial production in the original innovative design stage is not sufficient. It is necessary to develop dynamic measurement methods of spatiotemporal information for complex technical systems and complex production environments, such as the layout of unmanned production workshops and the measurement of processing timing information, and explore basic knowledge such as the flow of information at the minimal scale, the manufacturing physical process, and energy conversion, to meet the requirements of related technologies for multi-dimensional parameter acquisition and the analysis of product time-varying laws.

From the perspective of knowledge phenomenology, the essence of this process is the generation of knowledge grounded in inner temporal consciousness. The transformation of the spatiotemporal representation framework involves mapping original knowledge and data information through mathematical and structural transformations, thereby transferring them to new contexts. Leveraging the cognitive trinity of "past-present-future" in inner temporal consciousness, data from diverse spatiotemporal scenarios such as markets and industrial production are integrated to construct a technology-oriented cross-spatiotemporal cognitive system. This transferred knowledge is intricately intertwined with the requirements of technical tasks, forming a technical problem space for engineering practice. This space remains predominantly characterized by tacit knowledge, exemplified by numerous uncertain and difficult-to-explain dynamic parameters that contain substantial amounts of spatiotemporal-based tacit knowledge. However, due to digital representation enabled by dynamic measurement, the proportion of explicit knowledge within this space has significantly increased compared to the innovation solution space, providing parametric new knowledge that optimizes the original design scheme and facilitates the integration of explicit and tacit knowledge within the system.

5.1.2 Prediction of Technological Knowledge Base Based on Behavior-Structure Decoupling

The FBS design model accomplishes the sequential hierarchical mapping from product functions to behaviors and then to structures through eight steps, including conception, synthesis, analysis, evaluation, and description. Notably, behavior and structure provide detailed descriptions of the physical characteristics of the product, which are closely intertwined. These elements serve as the most vivid and intuitive cognitive representations of the innovation solution space and form the foundational knowledge framework for technology forecasting.

Specifically, the prediction of the technical knowledge base can be categorized into the following three aspects. ① Behavioral decoupling. Product behavior encompasses attributes derived from the design object's structure, indicating actions that the design object can perform through its established structure, such as the reciprocating motion of a hydraulic cylinder. Knowledge related to product behavior can be divided into two categories. The first category is based on traditional physical laws, emphasizing the kinematics and mechanics principles governing the product's structure. The second category comprises human expertise acquired through HRST in the original implicit knowledge space, often indirectly represented by complex mathematical and physical equations with numerous parameters. By implementing semantic-based refined classification for the first type of physical knowledge and conducting interpretable analysis of the parameters and equations involved in the second type of expertise knowledge, the knowledge embedded in product behavior can be quantitatively characterized. ②Structural decoupling. The product structure refers to the specific components of the design object, such as the cylinder barrel, cylinder head, piston and other components and dimensions of the hydraulic cylinder. In the design process, transforming or arranging physical effect carriers, screening of original principles, component configuration, which still belong to the category of qualitative design, and the relevant knowledge still remains at the conceptual expression level[49]. It is necessary to determine the product structure parameters through experimental research, simulation and precise calculation, strengthen the mathematical and physical representation, and support the subsequent intervention of quantitative technology. ③Comprehensive retrieval. Based on the knowledge spatiotemporal representation framework, the parametric expression of multi-granularity knowledge of behavior/structure is unified, and the retrieval based on design knowledge is carried out for the engineering technologies that meet the conditions to determine the mainstream technical solutions for the implementation of the key component functions.

From the perspective of knowledge phenomenology, this process represents the evolution of knowledge from the implicit to the explicit. The technical problem space encompasses a substantial amount of tacit knowledge, primarily derived from the learning and imitation of human skills within the original tacit knowledge domain, as well as product behavior structures that have not been fully parameterized during the design process. By decoupling behavior structures and applying formal synthesis, the technical problem space is transformed into quantifiable knowledge, information, and detailed technical constraints, thereby establishing an initial solution to the technical problem, refining the original design, and facilitating the conversion of tacit knowledge into explicit knowledge. Furthermore, the explicit knowledge generated in this process can be horizontally integrated with the original tacit knowledge domain to address inconsistencies in product function expansion.

5.1.3 Prediction of Technical Application Scenarios Based on Product Function Decoupling

In the FBS design model, product functions are the teleological descriptions of the product itself, indicating the intentions and purposes that the design object can achieve. For example, a hydraulic cylinder can realize the conversion of hydraulic energy to mechanical energy within the hydraulic system, as well as be applied to related engineering tasks and generate economic value and so on. From the perspective of the technical system, the function is the qualitative and quantitative description of the causal relationship between physical quantities, which can be understood as the state space for transforming inputs such as substances, energy, and signals, thereby forming the output that conforms to the design intention[49]. In the design process, the product function inspires the initial conception of the expected behavior and modifies the state space according to the adjustment of the behavioral structure in the subsequent design, forming a closed-loop self-optimization feedback. As a collection of series of state spaces, the product function plans the numerical range of each physical quantity within the technical system from the temporal and spatial dimensions and determines the technical application scenarios.

Specifically, the prediction of technology application scenarios can be divided into the following two aspects. ①Function decoupling. The knowledge involved in the iconic product functions includes three categories. The first is natural knowledge, referring to the objective physical information knowledge involved in the process of function realization. The second is social knowledge, referring to the social information and value involved in the functional service objects and application scenarios, such as the market that the product is targeted at and the characteristics of the user group. The third is thinking knowledge. Iconic products have human advanced intelligent characteristics, mainly derived from the learning and imitation of human thinking processes and skill behaviors. Thus, the types of knowledge involved in product functions are relatively complex, especially in the latter two types of knowledge, there is a large amount of tacit knowledge that cannot be completely quantified and represented, and it is necessary to further develop the human-machine collaborative comprehensive integration method. First, based on transplantable general technologies such as data mining and knowledge mining, explicit knowledge is extracted. Furthermore, based on expert experience judgment, combined with technologies such as human memory mechanism simulation and creative thinking toolboxes, potential tacit knowledge is analyzed. Finally, based on the giant knowledge base technology, the extracted knowledge is flexibly stored to enhance transferability[44]. ②Scene prediction. Based on the knowledge obtained from the decoupling of product functions, combined with technologies such as multi-modal intelligent perception, digital twin, and high-fidelity 3D modeling, establish the basic usage scenarios and production and manufacturing scenarios of digital products, calibrate the key coordinate positions online, and map the corresponding spatiotemporal structure of the scenarios. Further, integrate divergent thinking and convergent thinking, combined with the knowledge obtained from the decoupling of behavioral structures, restore the relevant human skills application scenarios, establish an open scene expression model with cross-mapping of knowledge content and spatiotemporal structure, and expand possible usage/manufacturing scenarios. Based on the requirements of existing technological solutions, conduct reliability screening for the above scenarios to ensure the feasibility of technology implementation.

From the perspective of knowledge phenomenology, this process represents the evolution of knowledge from explicit to implicit. By organically integrating the knowledge derived from FBS design decoupling, a spatiotemporal framework is constructed that facilitates active interaction between technologies and scenarios, enabling autonomous and controllable recording of behavioral data. During this interaction, the system autonomously emulates human skills and iteratively optimizes with behavioral data, thereby acquiring expertise and stimulating more disruptive technologies characterized by advanced embodied intelligence and deeper implicit knowledge. Furthermore, the implicit knowledge gained can be horizontally integrated into the original explicit knowledge domain, refining and correcting product function definitions according to the characteristics of the technological spatiotemporal scene. This ensures that the design meets implementation requirements, enhances the rationality of the design scheme, and paves the way for automated design processes.

5.1.4 Disruptive Technological Innovation Prediction Driven by Knowledge/Scenario Coupling

Based on the above knowledge base and application scenarios, develop disruptive technological innovation predictions driven by the coupling of knowledge and scenarios. Specifically, it can be divided into the following three aspects. ①Technology prediction driven by knowledge and scenarios. This process starts from the knowledge base, analyzes the spatiotemporal topological characteristics of the knowledge structure, establishes an association mapping mechanism from knowledge to scenarios, screens reasonable spatiotemporal scenarios, and ensures the identifiability and interpretability of the knowledge mapping. Further, technology can be understood as the intermediate interface and effect function that supports the input/output transformation of the system. According to the knowledge content, analyze the internal resources of the scenarios, identify the energy flow, material flow and information flow of the system input, correspond to the output functional goals, and explore the organizational structure and operation process that support the causal transformation of input/output. Combined with the knowledge base and the results of technology prediction, explore the possibility of generating new application scenarios. ② Scenario/Knowledge-driven technology forecasting. This process starts from the application scenarios, screens the knowledge that meets the characteristics of the scenarios and the functional requirements of the technology, and strives to represent the spatiotemporal structure of the technology application with the minimum knowledge scale and manage the technical knowledge in a lean manner. According to the spatiotemporal characteristics of the application scenarios and combined with the associated knowledge for technology realization, such as physical vector scaling and motion trajectory planning, the feasible range of technological activities is defined. Combined with the application scenarios and the results of technology forecasting, new knowledge that is suitable for technological activities is explored. ③ Comprehensive technology forecasting. Knowledge/Scenario-driven determines the basic technological structure and process, and Scenario/Knowledge-driven determines the feasible movement range of the technology. Through the dynamic and static coupling of the two, a multi-scale disruptive technology innovation forecasting system is established. On the other hand, new knowledge and new scenarios are organically integrated to expand the possibility space of technology forecasting and dynamically update the forecasting results.

From the perspective of knowledge phenomenology, this process fundamentally represents the integration phase of explicit and implicit knowledge evolution. On one hand, FBS design decoupling generates a substantial amount of explicit knowledge, thereby establishing a foundational framework for technological forecasting. On the other hand, scenario prediction transforms certain explicit knowledge into context-dependent implicit knowledge over time and space, while acquiring specialized knowledge by emulating human skills and behaviors in learning scenarios. Through deep cross-integration of knowledge and contexts, a comprehensive knowledge system for predicting disruptive technological innovation is established.

Further, iconic products need to evolve dynamically according to the cutting-edge strategic demands and adapt independently to the future industrial competitive situation. Based on the spatiotemporal characteristics of the connotation information of disruptive technologies, combined with technologies such as digital twin, virtual reality, and simulation, analyze the application and manufacturing scenarios of the new generation of iconic products, comprehensively predict the future market and technological competition focus, transform the technology prediction results into a technology resource pool, establish new design requirements and constraints, and implement the "engineering – design" task transformation based on spatiotemporal migration to support a new round of forward innovative design.

**5.2 Global Future Industry Competitive Landscape Based on Iconic Product Innovation**

Based on the systematic engineering of innovative iconic products, three core elements of knowledge, technology and products are extracted: Knowledge is the internal philosophical mechanism that drives the innovation of technology and products; Technology provides reliability guarantee for the implementation of products; Products are the comprehensive integration of technology and the ultimate value embodiment of the systematic engineering. Through the analysis of the scientific competitive situation, the development trend of frontier knowledge is grasped. Combined with the future technological competitive situation, technology prediction is elevated to the industrial dimension, and the global future industrial competitive situation is comprehensively analyzed.

Analysis of the scientific competition situation. Science, in essence, is a social existence and a complex and evolving giant system, which has different manifestations in different historical stages[44]. At present, the competitive situation of global scientific research can be summarized as follows. ① Basic research, as the source of the modern scientific system, through the exploration and discovery of the essence of the universe, life and matter, expands the cognitive territory of human beings. It has the characteristics of large resource investment and long return cycle, but it is crucial for the breakthrough development of human society and is generally regarded as a national strategy by developed countries. ②Interdisciplinary research reflects the comprehensiveness and complexity of the modern scientific knowledge system and is conducive to generating new frontier scientific knowledge[50]. It is the commanding height and main battlefield of international scientific competition. ③ Intelligent scientific research (AI4S), propelled by emerging artificial intelligence technologies, effectively tackles highly complex combinatorial explosion problems and realizes the intellectualization of the scientific research process as well as the automation of knowledge production [51]. This field represents a cutting-edge focus in current scientific competition. It is imperative to uphold the dialectical unity of reductionism and systems theory, enhance the accumulation of knowledge in basic research, promote interdisciplinary research across multiple domains, actively integrate advanced AI technologies, improve the efficiency of scientific research and knowledge production, and thereby gain strategic advantages in international scientific competition.

Analysis of the technological competition situation. Based on the innovative system engineering of iconic products, analyze the corresponding cutting-edge engineering technologies. ① Spatiotemporal big data, essentially a complex data set containing spatiotemporal coordinate information and having a multi-layer nested structure, through intelligent analysis, perceives and predicts the dynamic evolution spatiotemporal scenarios of event objects and establishes a three-dimensional cognitive framework. It is regarded as the key to informatization construction by European and American countries[52]. ② Human-machine collaboration, fully integrating the advantages of human perceptual awareness and qualitative synthesis with the rational logic and efficient iteration of machines, to support high-quality comprehensive decision-making and intelligent invention and creation, is the key development direction of the new generation of artificial intelligence. ③ Embodied intelligence, as a iconic product innovation enabling technology, through reinforcement learning technology, training multi-agent in complex environments and imitating human skills and behaviors, significantly improves the autonomous intelligence level of products and is the key to promoting the transformation of the physical world by artificial intelligence. For strategic high-tech fields, based on the comprehensive situation of science, technology, economy and social development, elevate technology forecasting to systematic technology foresight, select disruptive technologies that are expected to have a revolutionary impact on industrial development, and scientifically formulate technology roadmaps and resource allocation plans[53].

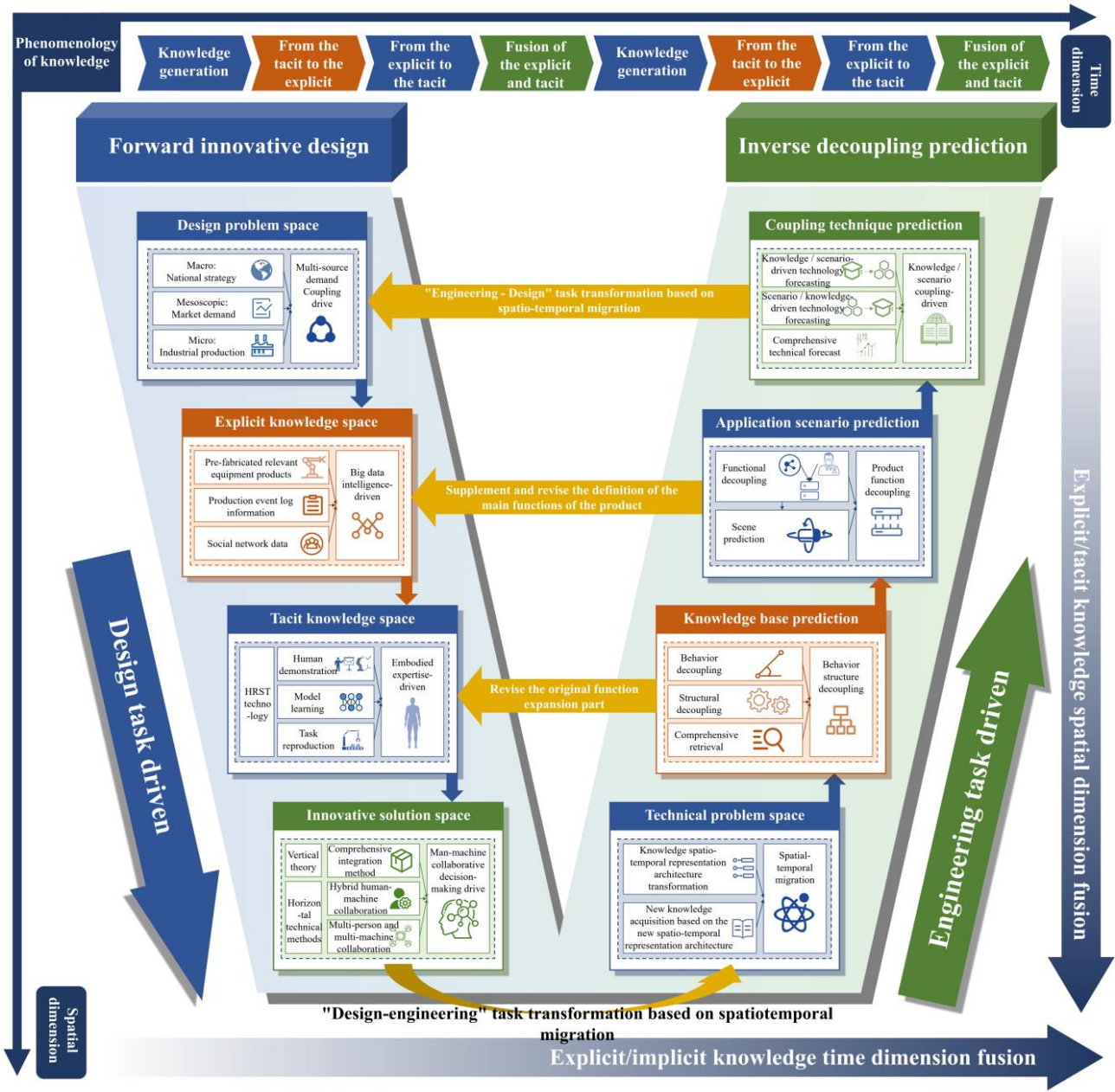

Figure 2　Iconic Product Innovation System Engineering

Analysis of industrial competition situation. Integrating scientific competition and technological competition, the following future industrial competition situations can be summarized. ①Jointly driven by time and space. In the time dimension, there are significant uncertainties and process risks in the development of future industries. All countries are stepping up to predict and develop disruptive technologies to seize strategic opportunities. In the spatial dimension, the development of future industries involves multiple factors such as international competition, scientific and technological level, economic activity, and regional spatial layout. The information value contained in spatiotemporal big data has increased unprecedentedly. ② Complete knowledge-driven. As the main carrier of future industries, the innovation of iconic products highly relies on the complete knowledge system that integrates explicit and tacit knowledge organically. On the one hand, complete knowledge endows products with human-like advanced intelligent features such as self-organization and self-adaptation, significantly enhancing the market competitiveness of products. On the other hand, as intangible capital and production factors, complete knowledge is an important strategic resource for the development of the knowledge economy and is regarded as the key to establishing and improving the national innovation system. ③System coupling-driven. Future industries are multi-level complex systems integrating science, technology, engineering, and the market, coupling multiple elements such as the demand side, supply side, policy side, and user side, involving multiple relationships among technology, products, organizations, and the environment, and having characteristics such as mutability, sensitivity, and emergence. It is necessary to adhere to the scientific thought of complex giant systems, based on a global and long-term perspective, support the exploration of frontier science and technology and the incubation of technology landing, lead the transformation of the social and economic development system, and promote the development of future industries to achieve a leap from quantitative change to qualitative change.

The aforementioned analysis of the competitive landscape of "science-technology-industry" adheres to the logic of knowledge evolution grounded in the knowledge Phenomenology. Scientific research, by uncovering the intrinsic laws of natural phenomena, transforms human perceptual judgments about the world into rigorous theoretical knowledge, marking a transition from latent to explicit knowledge. Technology then applies this explicit scientific knowledge to specific production and life processes, internalizing it through repeated practical applications into part of human experience, thereby facilitating a shift from explicit back to latent knowledge. Industry integrates science and technology seamlessly, establishing a sophisticated system for the national economy and industrial production, representing the synthesis of both explicit and latent knowledge.

## 5.3 Embodied Intelligence Enhancement Strategy for Iconic Products

Knowledge phenomenology is the logical core that supports the innovation of iconic products and the development of future industries. By establishing a complete knowledge system and comprehensively integrating the two thinking modes of "qualitative" and "quantitative", knowledge is elevated from the cognitive dimension to the wisdom dimension and transferred to the corresponding time and space. Therefore, based on knowledge phenomenology and HRST technology, this paper proposes the embodied intelligent enhancement strategy for iconic products (Figure 3), providing a technical method reference for the intelligent innovation of products. Specifically, it can be divided into the following four steps.

Top-down. This process aligns with the HRST teaching phase, where humans demonstrate to equipment products the skills and behaviors required for task execution. Through methods such as somatosensory camera capture and remote operation, movement trajectory information—including position, force, and velocity—is meticulously recorded. Additionally, brain-computer interfaces enable direct control of equipment for real-time behavior recording. This process can be conceptualized as knowledge generation within the framework of knowledge phenomenology. Through HRST teaching, the tacit knowledge embedded in human skills and behaviors is systematically mapped from physical space-time to digital space-time and transferred top-down to machines to establish a trainable dataset. This dataset can be considered an indirect representation of tacit knowledge, rich in information but with limited interpretability. Moreover, this process has not yet decomposed complex skills into discrete components; instead, it captures skills and behaviors as a continuous whole within a unified time sequence, consistent with the tripartite cognitive architecture of inner time consciousness.

Bottom-up. This process corresponds to the learning stage of the HRST model. By leveraging the behavioral characteristics of human skills, a dynamic model that aligns with the product structure is constructed. The aforementioned dataset is utilized for fitting and training to estimate model parameters and conduct interpretability analysis. This process can be understood as the transition from implicit to explicit knowledge in phenomenology. Through the establishment of interpretable models and parameters, and based on the explicit knowledge derived from existing theoretical models, the implicit knowledge embedded in the original data is mapped and transformed to fully explain the bottom-up movement process of skill behaviors.

Application and practice. This process

corresponds to the reproduction stage of the HRST task. The learned dynamic model is applied to the physical structure and controller of the product. Through control mechanisms such as position, speed, and torque, the product can replicate human skills, perform related tasks competently, and enhance proficiency through repeated iterations. This process can be understood as the transformation from explicit to implicit knowledge in phenomenology. Through the reproduction of actual tasks and iterative operations involving embodiment and contextualization in practice, the explicit knowledge acquired by the product through data training is internalized as implicit knowledge grounded in skill behaviors.

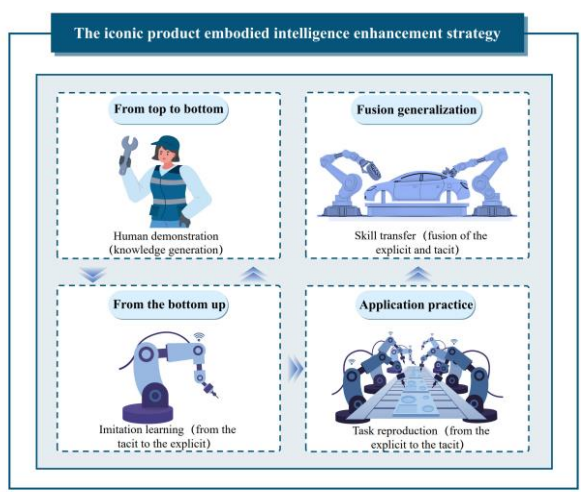

Figure 3 Embodied Intelligence Enhancement Strategy for Iconic Products

Fusion and Generalization. Leveraging the skills already acquired by the product, efforts are made to apply these skills across diverse task scenarios. By integrating multi-sensor and multi-modal information, the system predicts and perceives human intentions as well as novel scene characteristics. Skill transfer is facilitated through fine-tuning of the parameter models, thereby enhancing and generalizing the product's skill set and improving its environmental adaptability and intelligent operational capabilities. This process can be conceptualized as the fusion of explicit and implicit knowledge phenomenology. On one hand, skill generalization relies on the reuse of fused explicit and implicit knowledge from existing skills. On the other hand, cognitive learning in new scenarios involves establishing a comprehensive knowledge framework to enable the product to adapt to new spatiotemporal environments.

# 6 Conclusion

Iconic products are an important support for the development of future industries, the starting point and foothold of economic games and technological competitions among major countries, and the key to stimulating the national innovation potential, increasing social material and spiritual wealth, and improving the quality of life of the citizens. This paper integrates phenomenology, philosophy of technology and systems science to accurately identify the inherent logical differences between traditional products and iconic products; based on the essential reduction of phenomenology, it establishes the first principle of knowledge phenomenology of "knowledge generation - from implicit to explicit - from explicit to implicit - integration of explicit and implicit;" based on knowledge phenomenology and the law of human product design, it establishes the "design problem space - explicit knowledge space - tacit knowledge space - innovation solution space" forward innovative design framework for iconic products; combined with the FBS design theory, it reverse decouples the innovative design results and establishes the "technical problem space - knowledge base prediction - application scene prediction - coupling technology prediction" disruptive technological innovation prediction framework; by integrating forward innovative design and reverse decoupling prediction, it establishes the innovative system engineering of iconic products; through a comprehensive analysis of the global future industrial competition situation, it proposes the embodied intelligence enhancement strategy, providing a useful reference for the innovation research of iconic products.

To further promote independent innovation of iconic products, the following measures can be taken. ①Establish and improve the forward innovative design methods for iconic products in subdivided fields. Currently, forward innovative design is still at the theoretical stage, with insufficient consideration of specific technical details and design constraints. Expert teams in subdivided fields such as humanoid robots, brain-computer interfaces, and advanced and efficient aviation equipment can be organized. Combining the technical design requirements and knowledge distribution in each field, corresponding design methods can be studied and design manuals for subdivided fields can be compiled. ②Promote the innovation prediction of disruptive technologies in key fields and formulate technology roadmaps. Based on the six development directions of future industries and combined with various demands of social and economic development, led by government departments, establish a technology foresight platform covering universities, research institutions and enterprises, carry out national technology foresight activities, and comprehensively predict disruptive technological innovations; According to national strategies and market demands, formulate long-term macro technology roadmaps for the overall national competitiveness and short-term professional technology roadmaps for market implementation respectively. ③Strengthen the industrial application of disruptive technologies and iconic products. Establish a number of key laboratories, technology research centers and industrial development bases for the innovation of iconic products, improve supporting facilities and public service platforms, achieve efficient collaboration of "basic research and development - application implementation," and accelerate the industrialization process of high-tech and advanced products.

**Declaration of Conflict of Interest**

The author of this article hereby declares that there are no conflicts of interest or financial conflicts between them.

**References**

[1] Zhou J. Artificial Intelligence enables new industrialization. China Industry and Information Technology, 2024 (05): 38-43.
[2] Lu Y X. Thoughts on innovation and creation in the New era [J]. Science and Technology Review, 2024, 42(08): 1-2.
[3] Literature Research Office of the CPC Central Committee. Excerpts from Xi Jinping's discourse on scientific and technological innovation [M]. Beijing: Central Literary Publishing House, 2016.
[4] Zhou B, Leng F H, Li H, et al. Future industrial development deployment and enlightenment of major countries in the world [J]. Proceedings of the Chinese Academy of Sciences, 2021, 36(11): 1337-1347.
[5] Seven departments, including the Ministry of Industry and Information Technology, the Ministry of Education, and the Ministry of Science and Technology, jointly issued the "Implementation Opinions on Promoting Future Industrial Innovation and Development" [J]. Paper Making Information, 2024, (02): 6.
[6] Cusumano M A, Kahl S J, Suarez F F. Services, industry evolution, and the competitive strategies of product firms [J]. Strategic management journal, 2015, 36(4): 559-575.
[7] Tadevosyan R. Innovations and international competitiveness: country-level evidence [J]. Economics and Sociology, 2023, 16(3): 249-260.
[8] Melander L. Achieving sustainable development by collaborating in green product innovation [J]. Business strategy and the environment, 2017, 26(8): 1095-1109.
[9] Anderson A, McAllister C, Harris E. Product Development and Management Body of Knowledge: A Guidebook for Product Innovation Training and Certification [M]. John Wiley & Sons, 2024.
[10] Gaubinger K, Rabl M, Swan S, et al. Innovation and product management [J]. Innovation and product management: A holistic and practical approach to uncertainty reduction, 2015: 83-113.
[11] Wu Y, Zhou F, Kong J. Innovative design approach for product design based on TRIZ, AD, fuzzy and Grey relational analysis [J]. Computers & Industrial Engineering, 2020, 140: 106276.
[12] Russo D, Spreafico C. Investigating the multilevel logic in design solutions: A Function Behaviour Structure (FBS) analysis [J]. International Journal on Interactive Design and Manufacturing (IJIDeM), 2023, 17(4): 1789-1805.
[13] Suh N P. Axiomatic design theory for systems [J]. Research in engineering design, 1998, 10: 189-209.
[14] Li J, Guo X, Zhang K, et al. A knowledge-enabled approach for user experience-driven product improvement at the conceptual design stage [J]. AI EDAM, 2023, 37: e22.
[15] Xu J, Tao M, Gao M, et al. Assembly precision design for parallel robotic mechanism based on uncertain hybrid tolerance allocation [J]. Robotic Intelligence and Automation, 2023, 43(1): 23-34.
[16] Zhou J, Zhou Y, Wang B, et al. Human–cyber–physical systems (HCPSs) in the context of new-generation intelligent manufacturing[J]. Engineering, 2019, 5(4): 624-636.
[17] Yao X F, Zhou J J. Theory and Technology of Intelligent Manufacturing [M]. Beijing: Science Press, 2020.
[18] Zhou W, Xu L Y. On new quality productivity: connotation, characteristics and important focus [J]. Reformation, 2023(10): 1-13.
[19] Qu H X, Xu J, Xu J Y. Spatiotemporal Phenomenology of Disruptive Technological Innovations in Future Industries [J].

Strategic Study of CAE, 2024, 26(03): 239-256.

[20] Department of Engineering and Materials Science, National Natural Science Foundation of China. Mechanical engineering discipline development strategy report (2021—2035) [M]. Beijing: Science Press, 2021.

[21] Yao X F, Huang Y, Huang Y S, et al. Autonomous smart manufacturing: Social-cyber-physical interaction, reference architecture and operation mechanism [J]. Computer Integrated Manufacturing Systems, 2022, 28(2): 325–338.

[22] Li J H, Fu Z, Zhang J. Computationalism: A New World View [M]. Beijing: China Social Sciences Press, 2012.

[23] Polanyi M. The logic of tacit inference [J]. Philosophy, 1966, 41(155): 1–18.

[24] Russell. Human Knowledge [M]. Beijing: The Commercial Press, 1983.

[25] Wang Z T. Knowledge Management [M]. Beijing: Science Press, 2009.

[26] Wang Z T. Knowledge systems engineering [M]. Beijing: Science Press, 2009.

[27] Dan Zahavi. Husserl's Phenomenology [M]. Beijing: The Commercial Press, 2022.

[28] Husserl. Phenomenology of Inner time Consciousness [M]. Beijing: The Commercial Press, 2009.

[29] Dreyfus H. How far is distance learning from education? [J]. Bulletin of Science, Technology & Society, 2001, 21(3): 165-174.

[30] Merleau-Ponty M. Phenomenology of perception [M]. London: Routledge, 2002.

[31] Selinger E, Crease R P. The philosophy of expertise [M]. Columbia University Press, 2006.

[32] Dreyfus H L, Dreyfus S E. What artificial experts can and cannot do [J]. AI & society, 1992, 6: 18-26.

[33] Tan J R. Key technologies and development trends of innovative design [J]. Enterprise Management, 2023(05): 11-12.

[34] Tan J R, GAO M Y, Xu J H, et al. Digital Intelligent forward Design Method and Its Application in Manufacturing Equipment and Process [J]. Chinese Journal of Mechanical Engineering, 2023, 59(19): 111-125.

[35] Innovative Design Development Strategy Research Team. China Innovation Design Roadmap [M]. Beijing: Science and Technology Press of China, 2016.

[36] Lu Y X, Sun S Q, Zhang K J. Research on Development strategy of Innovative Design [J]. Machine Design, 2019, 36(02): 1-4.

[37] Lu Y X. On Innovative Design [M]. Beijing: Science and Technology Press of China, 2017.

[38] Yang L C, Li H, Li B. Product Design Knowledge Management System and its Application [M]. Beijing: Science Press, 2009.

[39] Liu A. Problem definition[J]. Design Engineering and Science, 2021: 167-189.

[40] Goldschmidt G. Capturing indeterminism: representation in the design problem space[J]. Design Studies, 1997, 18(4): 441-455.

[41] Fu J P, Zhao H Y, Cao J, et al. Anomaly detection algorithm for business process control flow based on event log: Status and evaluation [J]. Computer Integrated Manufacturing Systems, 2024, 30(08): 2631-2643.

[42] Zeng C, Yang C G, Li Q, et al. Research progress of Human-robot skill transfer [J]. Acta Automatica Sinica, 2019, 45(10): 1813-1828.

[43] Wu X D, Wang X F, Jin B, et al. Human-machine Synergy [J]. Beijing: Science Press, 2022.

[44] Zhao S K. Exploration of the overall framework of modern science and technology system [M]. Beijing: Science Press, 2011.

[45] Huang S J. Marxist philosophy and modern science and technology system [M]. Beijing: Science Press, 2011.

[46] Gero J S. Design prototypes: a knowledge representation schema for design [J]. AI magazine, 1990, 11(4): 26-26.

[47] Forsberg K, Mooz H. The relationship of system engineering to the project cycle [J]. Center for Systems Management, 1991, 5333.

[48] Li H F, Zhu J W. The "shape" of spatiotemporal big Data: a geometric and topological perspective [M]. Beijing: Science Press, 2024.

[49] Koller R. Mechanical design methodology [M]. Beijing: Science Press, 1990.

[50] Lu Y X. The significance of interdisciplinary and interdisciplinary science [J]. Proceedings of the Chinese Academy of Sciences, 2005, (01): 58-60.

[51] Li G J. Intelligent Research (AI4R) : the fifth research paradigm [J]. Proceedings of the Chinese Academy of Sciences, 2024, 39(01): 1-9.

[52] Wu X C. Spatial-Temporal big data and cloud platform (Theory) [M]. Beijing: Science Press, 2018.

[53] "China Engineering Science and Technology 2035 Development Strategy Research" Project Team. The development strategy of China's engineering science and technology for 2035 [M]. Beijing: Science Press, 2019.